\documentclass[reprint,amsmath,superscriptaddress,amssymb,aps,prl,showpacs,floatfix]{revtex4-1}

\usepackage{graphicx}
\usepackage{dcolumn}
\usepackage{bm}
\usepackage{hyperref}
\usepackage{subfigure}

\begin{document}


\title{Charge-noise tolerant exchange gates of singlet-triplet qubits in asymmetric double quantum dots}

\author{Tuukka Hiltunen}
\affiliation{%
COMP Centre of Excellence, Department of Applied Physics, Aalto University, Helsinki, Finland
}%

\author{Hendrik Bluhm}
\affiliation{JARA-Institute for Quantum Information, RWTH Aachen University, D-52056 Aachen, Germany}

\author{Sebastian Mehl}
\affiliation{Peter Gr\"unberg Institute (PGI-2), Forschungszentrum J\"ulich, D-52425 J\"ulich, Germany}
\affiliation{JARA-Institute for Quantum Information, RWTH Aachen University, D-52056 Aachen, Germany}

\author{Ari Harju}
\affiliation{%
COMP Centre of Excellence, Department of Applied Physics, Aalto University, Helsinki, Finland
}%

\date{\today}

\begin{abstract} 
In the semi-conductor double quantum dot singlet-triplet qubit architecture, the decoherence caused by the qubit's charge environment poses a serious obstacle in the way towards large scale quantum computing.
The effects of the charge decoherence can be mitigated by operating the qubit in the so called sweet spot regions where it is insensitive to electrical noise. In this paper, we propose
singlet-triplet qubits based on two quantum dots of different sizes. Such asymmetric double dot systems allow the implementation of exchange gates with controllable exchange splitting $J$ operated
in the doubly occupied charge region of the larger dot, where the qubit has high resilience to charge noise. In the larger dot, $J$ can be quenched to a value smaller than the intra-dot tunneling using magnetic fields, while the smaller dot
and its larger splitting can be used in the projective readout of the qubit.
\end{abstract}

\maketitle

The two-electron unpolarized singlet and triplet states in semi-conductor double quantum dots (DQD) are a promising scalable realization for a quantum bit \cite{levy,taylor}.
The universal set of qubit operations \cite{petta2, guido, foletti, shulman} in this architecture includes one-qubit rotations
generated by electrically detuning the two dots of the DQD-system.
These exchange rotations are dephased due to the charge noise caused by the electrical environment of the qubit \cite{hu1,weperen,petta2,charge_noise1,maune,dial,ramon}. Charge noise can be represented by voltage-noise in the detuning of the qubit \cite{dial}, which results in fluctuations in the exchange splitting $J$
that affect the frequency of the exchange rotations and cause decoherence. The charge-based decoherence is a severe factor that limits the performance of the singlet-triplet qubits. Thus,
there has been several proposals for mitigating its effects, including multi-electron singlet-triplet qubits \cite{screen,mehl1} and optimized gate sequences \cite{kestner,cerfontaine}.

Another widely investigated possibility is to exploit the so called sweet spot regions where the exchange splitting is insensitive to charge noise in the gate operations \cite{taylor2,requ,mehl1,dial}.
For example, in the far detuned region, where both the singlet and the triplet are in the doubly occupied charge states, the qubit is much less susceptible to charge noise as both of the
qubit states have similar charge densities \cite{dial}.
Utilizing this insensitive region for gate operations requires rapid switching between the sweet spot and the singly occupied configuration with one electron in each dot. 
To prevent excitation into higher orbital states during this transfer of one electron from one dot to another, 
the corresponding change in detuning needs to be adiabatic with respect to the tunnel coupling \cite{leak}. 
On the other hand, the phase accumulated in the doubly occupied configuration should be as small as possible (of order $\pi$) in order to minimize dephasing. 
Thus, the switching time should be on the order of $1/J$. Together with the adiabaticity requirement, this condition implies that the the tunnel coupling must be larger than the exchange splitting. 
Furthermore, the limited speed of control electronics favor switching times not much faster than 1 ns so that exchange splittings exceeding a few $\mu$eV are practically cumbersome.
On the other hand, singlet-triplet qubits typically employ Pauli blockade for readout via spin to charge conversion, 
which requires an exchange splitting larger than the tunnel coupling to maintain good charge contrast. 
Hence, one faces two conflicting requirements: small J for high fidelity gates, but large J for the readout.

\begin{figure}[!ht]
\vspace{0.3cm}
\includegraphics[width=0.9\columnwidth]{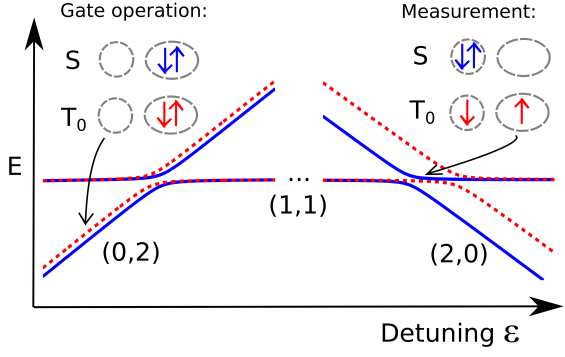}
\caption{(Color online) The ADQD scheme. The energies of the singlet (blue lines) and triplet (dashed red lines) are shown in the two dots [$(0,2)$ and $(2,0)$-configurations] as functions of the detuning of the dots $\epsilon$.
The exchange gate operations are done in the larger right dot so that both the singlet and triplet are in the $(0,2)$-configuration. The projective measurement is conducted using the smaller left dot, with the singlet in $(2,0)$ and the triplet in $(1,1)$. }
\label{fig:schem}
\end{figure}

In this paper, we propose an asymmetric double quantum dot (ADQD) system, consisting of two quantum dots with different sizes, that allows exchange-gate operations with high tolerance to
charge noise in the doubly occupied region of one dot while double occupation of the other dot is used for readout (see Fig. \ref{fig:schem}).
Using out of plane magnetic fields, the exchange splitting can be set to a small non-zero value for the exchange gate in the doubly occupied region of the larger dot. Due to the size difference, the exchange stays
large in the smaller dot which can be used in the projective readout of the system.
Thus, the conflicting requirements outlined above can be met simultaneously.
Note that asymmetric double dots have been proposed previously in the context of the so called inverted singlet-triplet qubits \cite{inverted}.

The DQD two-electron system (confined to the $xy$-plane) is described with the continuum Hamiltonian
\begin{equation}
\label{eq:Hamiltonian}
\hat{H}=\sum^2_{j=1}\left[\frac{(\hat{\mathbf{p}}_j+e\mathbf{A}(\mathbf{r}_j))^2} {2 m^*}+V(\mathbf{r}_j)\right] +\frac{e^2}{4\pi\epsilon r_{12}}.
\end{equation}
Here, $\mathbf{A}(\mathbf{r}_j)=\frac{1}{2}B_z(-y_j,x_j,0)$ is the magnetic vector potential corresponding to a homogeneous external magnetic field $B_z$ and $V$ the electric potential. $m^*\approx0.067\,m_e$ and
$\epsilon\approx12.7\,\epsilon_0$ are the effective electron mass and
permittivity in GaAs, respectively. 

The electric potential $V(\mathbf{r})=V(x,y)=V_{c}(x,y)+V_{d}(x,y)$ consists of the QD confinement $V_c$ and the detuning potential $V_d$. We model the DQD system
as two parabolical wells located at the $x$-axis at $\mathbf{R}_1=\left(-\frac{a}{2},0\right)$ and $\mathbf{R}_2=\left(\frac{a}{2},0\right)$, where $a$ is the distance of the QD minima.
The detuning $V_{d}(x,y)$ is modeled as a step function that assumes the value $-\frac{\epsilon}{2}$ in the left dot (the one at the negative $x$ axis) and the
value $\frac{\epsilon}{2}$ in the right one, with $\epsilon=V(\mathbf{R}_2)-V(\mathbf{R}_1)$ being the energy difference between the dots. 

\begin{figure}[!ht]
\vspace{0.3cm}
\includegraphics[width=\columnwidth]{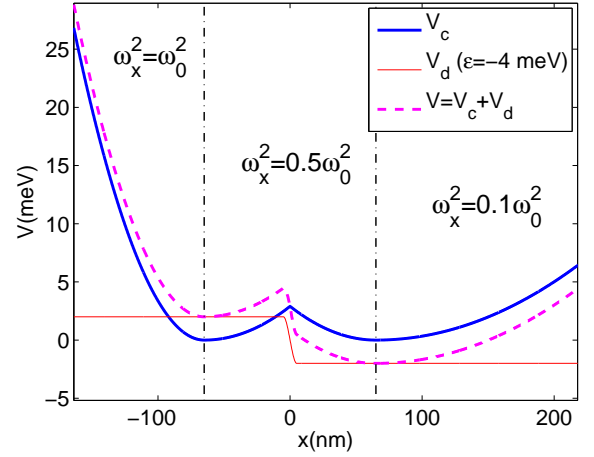}
\caption{(Color online) The electric potential in the $x$-axis of the DQD system. The piecewise parabolic confinement $V_c$ is shown as the blue line. The minima of the dots are at $x=\pm65$ nm on the $x$-axis and the confinement strength in the $y$-direction is $\hbar\omega_0=2.5$ meV.
The regions of different $x$-confinements $\omega_x$ are shown with the dashed vertical lines. The detuning potential $V_d$, with $\epsilon=-4$ meV is shown in red, and the combined electric potential $V=V_c+V_d$ as the dashed purple line.}
\label{fig:pot}
\end{figure}

The electric potential of the ADQD system is illustrated in Fig. \ref{fig:pot}. The potentials $V_c$, $V_d$, and $V=V_c+V_d$ are shown in the $x$-axis. The minima of the dots are $a=130$ nm apart, located on the $x$-axis, at $\pm65$ nm. The confinement is piecewise parabolical, meaning that the confinement strength
in the $x$-direction
has different values in different regions. 
The (singlet) intra-dot tunneling has the values $55$ $\mu$eV and $38$ $\mu$eV
at $B_z=0$ and $B_z=0.87$ T magnetic fields, respectively.
The actual form of the dots (e.g. whether they are elliptical or circular) was not found to have significant effects on the physics of the system. The piecewise quadratic form was chosen because it allows the control of the tunneling and the $J$-splitting independently.

\begin{figure}[!ht]
\vspace{0.3cm}
\includegraphics[width=\columnwidth]{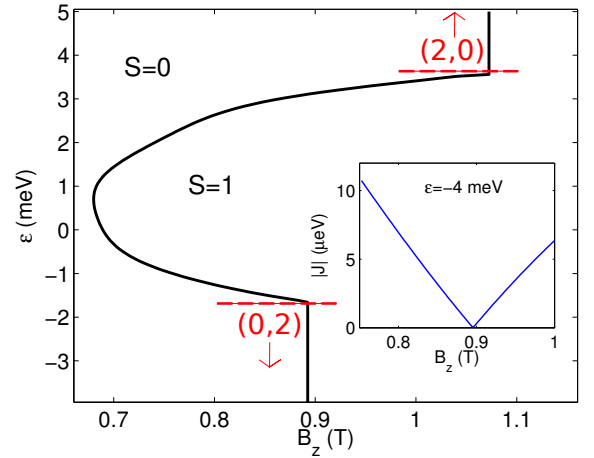}
\caption{The ground state spin as a function of the magnetic field $B_z$ and the detuning $\epsilon$ (negative values of $\epsilon$ correspond to the right dot in low potential and positive to the left dot) in the ADQD-system of Fig. \ref{fig:pot}.
The boundary curve between the $S=0$ and $S=1$ ground states is shown with the black line. The dashed red lines denote the boundaries of the regions where both the singlet and the triplet are in the doubly occupied charge states.
Inset: The magnetic field dependence of the absolute value of the exchange energy $J$ in the larger dot.
Here, the detuning is $\epsilon=-4$ meV, corresponding to the region where both $|S\rangle$ and $|T_0\rangle$ are in the $(2,0)$-charge configuration.}
\label{fig:mfield}
\end{figure}

The Hamiltonian (\ref{eq:Hamiltonian}) is diagonalized using the exact diagonalization (ED) method and the Lanczos algorithm.
We use 50 first single-particle orbitals to build the many-body basis. This is found to be sufficient for the convergence of
the results (see the supplementary material Ref. \onlinecite{suppl}).

In singlet-triplet qubits, the exchange interactions create an energy splitting $J=E_{T_0}-E_S$ between singlet $|S\rangle$ and $S_z=0$ triplet $|T_0\rangle$. When the qubit
is in the $(1,1)$-charge configuration (i.e. one electron in each dot), the exchange is typically very close to zero, in the sub $\mu$eV region. It can be turned on by detuning one of the dots of the DQD-system to lower potential.
As the detuning is increased, it will eventually overcome the Coulomb-repulsion and both electrons will localize to the dot with the lower potential. 
In the zero or low magnetic field, the $S=0$ singlet is the ground state and the transition to the doubly occupied $(2,0)$ and $(0,2)$-states happens at lower detuning values in the $|S\rangle$-state than in the $|T_0\rangle$-state.
However, increasing the $B_z$-magnetic field will eventually shift the triplet as the ground state \cite{harju}. The ground state spin in the DQD system of Fig \ref{fig:pot} is plotted in Fig. \ref{fig:mfield} as a function of the
magnetic field $B_z$ and the detuning $\epsilon$.

As seen in the figure, the spin phase boundary becomes a straight line at high detuning ($\epsilon<-2$ meV or $\epsilon>4$ meV), i.e. the transition to the $S=1$ ground state happens at a fixed value of $B_z$ regardless of the detuning.
This is due to the transition to the doubly occupied $(2,0)$ and $(0,2)$-states. When both the singlet and the triplet have undergone the transition, the system behaves
as a doubly occupied single dot, and the detuning just lowers the energies $E_S$ and $E_{T_0}$ but keeps $J$ (approximately) constant. In the figure,
the transition value is $B_z=0.893$ T in the right dot and $B_z=1.07$ T in the smaller left dot. 
The transition values of $B_z$ depend on the confinement
strengths of the dots. The larger the dot, the lower the transition value.
Along the transition boundary, the $|S\rangle$ and $|T_0\rangle$ states are degenerate. 

The ADQD system allows the implementation of single-qubit exchange gates that are operated in the doubly occupied region with a magnetically controllable value of $J$.
The perpendicular magnetic field is set to a value that is close to the $S=1$ transition in the larger dot to obtain a $J$ splitting of a few $\mu$eVs (smaller than the intra-dot tunneling) in the doubly occupied $(0,2)$-region of
the larger right dot. We found that in the zero magnetic field, one would need to have very large dots with wave function diameters close to 1 $\mu$m to quench $J$ this small in the doubly occupied region.

The inset in Fig. \ref{fig:mfield} shows the absolute value of the exchange energy in the doubly occupied region of the larger right dot as a function of the magnetic field $B_z$.
Here, the detuning is $\epsilon=-4$ meV, corresponding to the region where both the singlet and the triplet are
in the $(0,2)$-charge configuration. As seen in the figure, $J$ is approximately linear in $B_z$ close to the $S=1$ transition values. At the spin phase boundary
(at $B_z=0.893$ T), $J$ changes sign, i.e. the triplet becomes the ground state, as seen in the kink at the $|J|$-curve of the inset.

\begin{figure*}[!ht]
\vspace{0.3cm}
\includegraphics[width=\columnwidth]{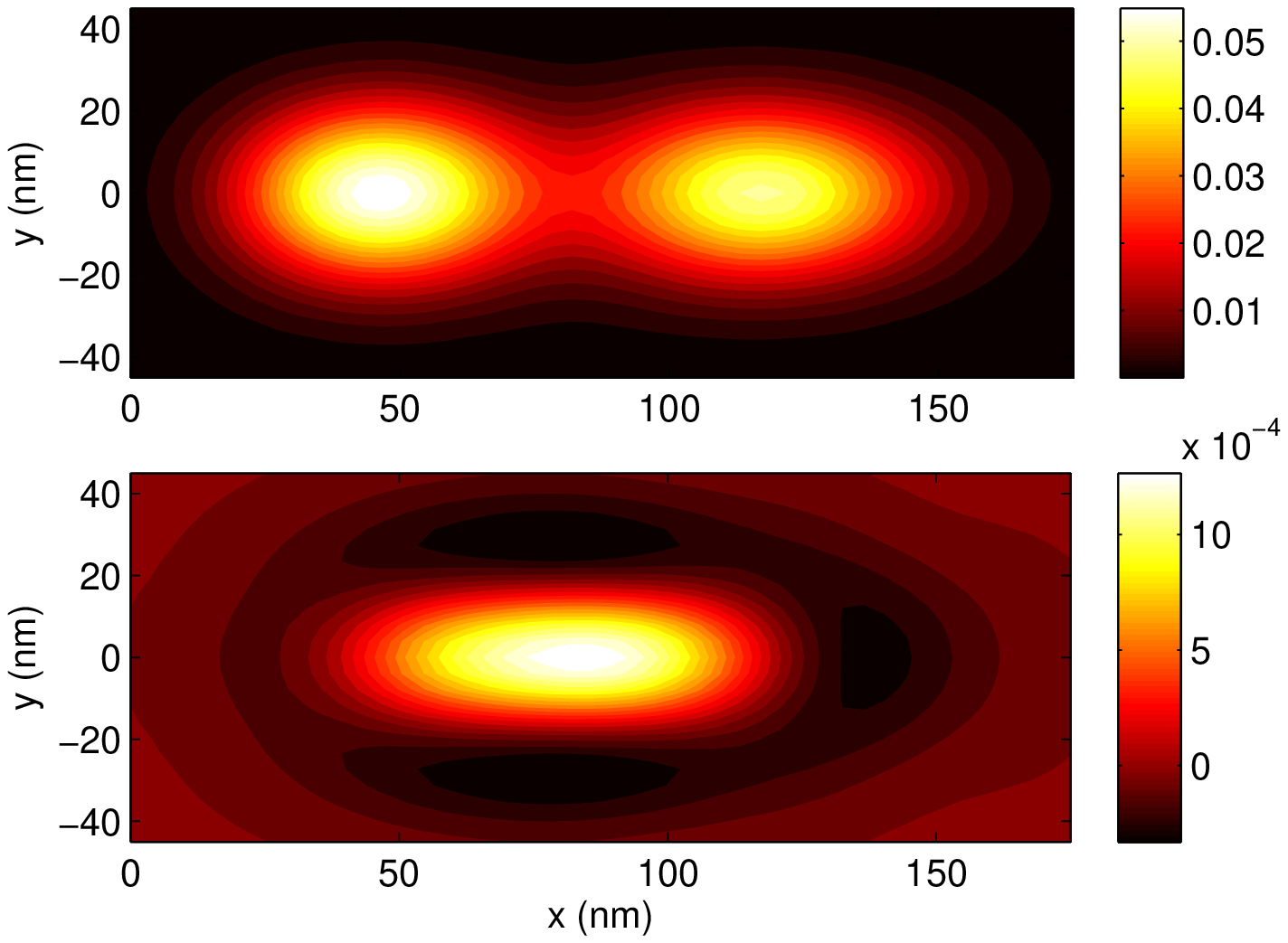}
\includegraphics[width=\columnwidth]{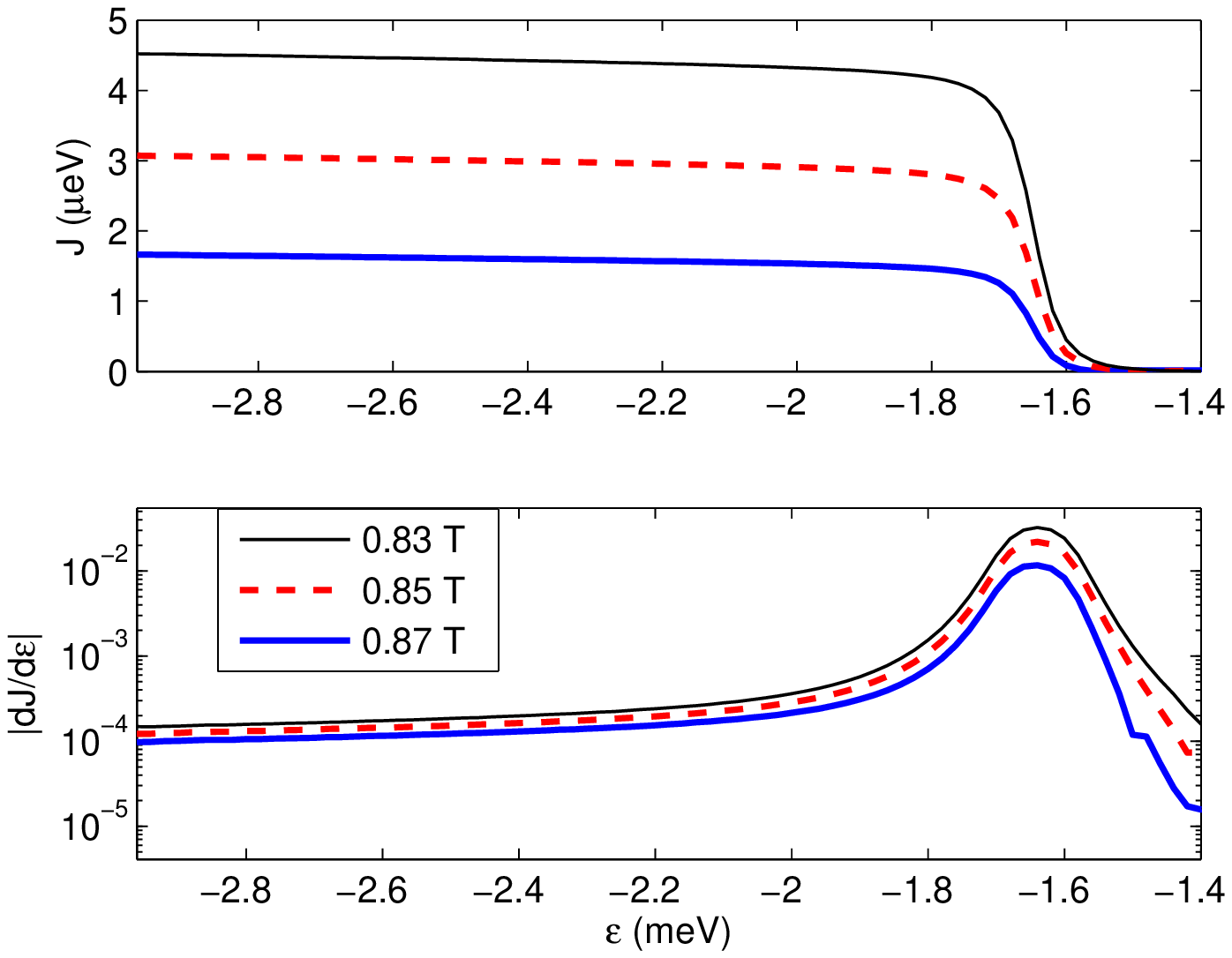}
\caption{(Color online) Left: the charge density of the singlet, $\rho_S$, (the upper plot) and the difference of the singlet and the triplet, $\rho_S-\rho_{T_0}$, (the lower plot) in the $(0,2)$-configuration with the magnetic field strength $B_z=0.87$ T (the states are localized in the right dot, so the left dot is omitted in the pictures). The ADQD-system parameters are as in Fig. \ref{fig:pot}. The detuning is $\epsilon=-4$ meV, corresponding to
the exchange energy of $J=1.7$ $\mu$eV. Here, the unit of the densities is ${e}$/nm$^2$. Right: The exchange energy $J$ (upper plot) and its derivative $\text{d}J/\text{d}\epsilon$ (lower plot) as functions of the detuning $\epsilon$. The results are shown with
several values of $B_z$ taken near the spin phase transition at $B_z=0.893$.}
\label{fig:exchange}
\end{figure*}

The $|S(0,2)\rangle$ and $|T_0(0,2)\rangle$ states have close to identical charge densities, as shown in the the left panels of Fig. \ref{fig:exchange}, 
allowing protection against electrical noise. The charge difference between the singlet and the triplet in the left dot is $\Delta q^{right}=q_{S}^{right}-q_{T_0}^{right}=4.7\times10^{-5}e$ (obtained by integrating the difference of the lower plot). This is in stark contrast to the traditional
exchange gate implementation, where the gate is operated near the singlet $(0,2)$-transition while the triplet stays fully in $(1,1)$. In this case, the charge difference (corresponding
to same $J$-value as in the doubly occupied case above) is $\Delta q^{right}=q_{S}^{right}-q_{T_0}^{right}=2.7\times10^{-2}e$, more than three orders of magnitude larger.
In the $(0,2)$-configuration, the qubit states are also protected from the hyperfine induced decoherence, as both electrons are localized in the same dot and the hyperfine effects are suppressed under exchange \cite{petta2}.

Decoherence by charge noise in $S$-$T_0$ qubits has been measured to behave as $\epsilon$-noise \cite{dial}, meaning that charge noise manifests itself as effective fluctuations in the qubit's detuning $\epsilon$.
The decoherence from charge noise is thus mainly governed by $|\text{d}J/\text{d}\epsilon|$. The right panels of Fig. \ref{fig:exchange}
show $J$ and its derivative $|\text{d}J/\text{d}\epsilon|$ as a function of the detuning $\epsilon$. As seen in the figures, $J$ stays approximately constant after the charge transitions ($\epsilon<-2$ meV).
There is however a small, close to linear, $\epsilon$-dependence even in the $(0,2)$-region. This is explained by the fact that the wave functions of the singlet and triplet have small finite values
in the barrier between the dots.

In any case, the values of $|\text{d}J/\text{d}\epsilon|$ in this proposed $(0,2)$-operation stay much lower than in the corresponding traditional $S(0,2)-T_0(1,1)$ exchange gates. In, for example, the
$B_z=0.850$ T curve that corresponds to $J=3.07$ $\mu$eV, the derivative has the value $|\text{d}J/\text{d}\epsilon|=1.26\times10^{-4}$. If one were to create the same exchange splitting $J=3.07$ $\mu$eV
in the $(1,1)-(0,2)$ transition region, the derivative would be several hundred times larger, $|\text{d}J/\text{d}\epsilon|=2.59\times10^{-2}$, corresponding to the case where the smaller dot in Fig. \ref{fig:pot} is detuned to low energy. 
In the bigger dot or nonzero fields, the derivative would be even larger, as the $S-T_0$ splitting stays smaller, and larger charge density differences are needed for the same exchange splitting.

The loss of coherence for a given pulse sequence, evolution time and noise spectrum, which is the ultimate figure of merit for qubit operation, can be shown to be proportional to $(\text{d}J/\text{d}\epsilon)^2$ \cite{charge_noise1,charge_noise2}. Thus, we find that an improvement of two orders of magnitude is possible. 
When considering dephasing times, $\text{d}J/\text{d}\epsilon$ enters linearly for $T_2^*$ arising from quasi-static noise, and quadratically for $T^*_2$ arising from white noise.
With realistic noise-parameters \cite{dial} and assuming quasi-static noise, the coherence time in the $J=3.07$ $\mu$eV operation is $T_2^*=6.2$ $\mu$s in the $(0,2)$-operation (see the supplementary material for details).
The coherence time for the same operation in the traditional $(1,1)$-regime would be (with the same noise parameters) $T^*_2=30$ ns (see the supplementary material).

\begin{figure}[!ht]
\vspace{0.3cm}
\includegraphics[width=\columnwidth]{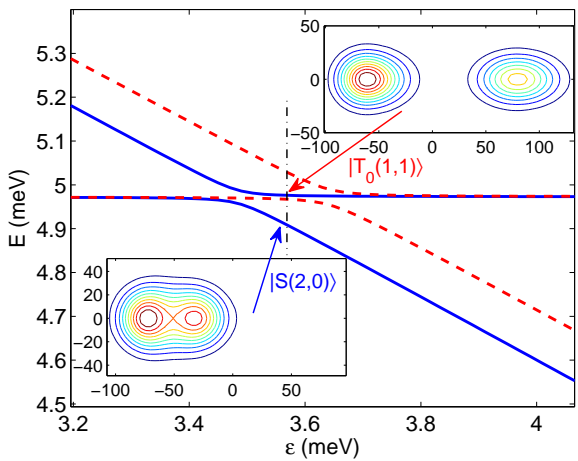}
\caption{(Color online) The energies of the lowest singlet and triplet states at their charge transition anti-crossings in the smaller left dot of the ADQD-system of Fig. \ref{fig:pot}. The magnetic field is
$B_z=0.87$ T, which corresponds to close to identical charge densities in the larger dot (see Fig. \ref{fig:exchange}).
The singlet states are shown with blue lines and the triplets with red dashed lines. The insets show the densities (the unit is ${e}$/nm$^2$) of the lowest $|S\rangle$ and $|T_0\rangle$-states taken at $\epsilon=3.57$ meV (denoted by the arrows and the dash-dotted vertical line in the figure).}
\label{fig:measure}
\end{figure}

Next, we discuss the projective readout \cite{petta2,taylor2} of the ADQD-system done in the smaller left dot. The singlet
probability is measured by sweeping the detuning to a region where only the singlet is in the doubly occupied configuration. The difference in the values of $B_z$ corresponding to the $S=1$
transition in $(2,0)$ and $(0,2)$ allows the smaller dot to have large $J$ values and charge density differences between the qubit states while in the right dot the doubly occupied states are nearly degenerate. The anti-crossing of the charge states in the left dot for the same system as in Figs. \ref{fig:mfield} and \ref{fig:exchange} is shown in Fig. \ref{fig:measure}.
Between the singlet and triplet anti-crossings ($3.45$ meV $<\epsilon< 3.70$ meV), there is a region where the singlet is fully in $(2,0)$ while the triplet is in $(1,1)$. The insets in the left panel show the singlet and triplet charge densities taken at
$\epsilon=3.57$ meV between the anti-crossings. In the insets, the amount of charge in the right dot is $q_{T_0}^{right}=0.900e$ in the triplet and $q^{right}_{S}=0.0993e$ in the singlet.

Finally, we will shortly discuss general gate-operation in ADQD singlet-triplet qubits. The $x$-rotations in the Bloch sphere are generated with magnetic field gradients like in the conventional $S-T_0$-qubits \cite{foletti}. 
The magnetic field gradient between the two dots of the system can be simulated by adding a Zeemann-term $V_Z(\mathbf{r})=g^*\mu_BB_{nuc}(\mathbf{r})S_z$ to the Hamiltonian of Eq. (\ref{eq:Hamiltonian}).
Here, the inhomogeneous magnetic field $B_{nuc}$ is modeled as a step function that assumes constant values at each dot. As expected, these additional simulations show that the $x$-rotations done in the $(1,1)$-charge configuration
are found to work completely similarly to the conventional case. The two-qubit operations can also be implemented the same way as conventionally: capacitatively (using the smaller dot and the 'typical' $S(0,2)-T_0(1,1)$ detuning regime) \cite{taylor,stepa,shulman} or with exchange based methods \cite{levy,li}. The
latter would benefit from the improved charge noise resilience discussed here.
There has also been a proposal for using the double occupation region in the capacitative coupling \cite{Nielsen}, which could offer large enhancements to the coherence times in the two qubit-operation.
This scheme could also benefit from the ADQD $S$-$T_0$ qubit implementation and the magnetic field control of $J$, as it would allow the quenching of $J$ to the sub-tunneling scale. 

A potential difficulty when implementing the ADQD scheme discussed in this paper is that since J is independent of detuning, the rotation angle of a gate cannot be controlled by the pulse amplitude. Instead, the pulse duration must be used, 
which is less flexibly controllable on current pulse generators. Furthermore, larger voltage pulses spanning all the way from (0, 2) to (2, 0) are required for readout.

In summary, we have simulated a singlet-triplet qubit based on an asymmetric double quantum dot system. The size difference of the dots allows the larger one to be used in exchange gate operations with a moderate and controllable $J$-splitting done in the far detuned $(0,2)$-regime, while when detuning the smaller dot to low potential, the
splitting stays large enough for the projective readout of the qubit. In the far $(0,2)$-regime, the $S$- and $T_0$-states have similar charge densities which results in weaker coupling between the qubit and its
charge environment. The detuning dependence of $J$ was found to be very small in the $(0,2)$-region, resulting in high resistance to $\epsilon$-noise, that is the dominant form of charge noise. The ADQD-scheme allows for a noise resistant implementation of exchange gates in singlet-triplet qubits,
alleviating the crucial problem of decoherence in this quantum computing architecture.

\begin{acknowledgments}
This research has been supported by the Academy of Finland through its Centres of Excellence Program (project no. 251748).
\end{acknowledgments}
\bibliography{lagrange}

\end{document}